\documentclass[pra,nofootinbib,amssymb,amsmath]{revtex4}

\usepackage{appendix}
\usepackage{graphicx}
\usepackage{amssymb,amsmath,amsbsy} 

\usepackage[polish,english]{babel}
\usepackage[T1]{fontenc}

\def\B#1{\left(#1\right)}
\def\BB#1{\left[#1\right]}

\def\be{\begin{equation}}
\def\ee{\end{equation}}

\def\bee{\begin{equation*}}
\def\eee{\end{equation*}}

\def\mn{g^2-2g\cos(k)+1}

\def\la{\langle}
\def\ra{\rangle}

\begin{document}

\title{Counterdiabatic driving of the quantum Ising model}

\author{Bogdan Damski}
\affiliation{Institute of Physics, Jagiellonian University, {\L}ojasiewicza 11, 30-348 Krak\'ow, Poland}

\begin{abstract}
The system undergoes adiabatic evolution when its population in the
instantaneous eigenbasis of its time-dependent Hamiltonian changes only negligibly.
Realization of such dynamics requires slow-enough  changes of the 
parameters of the Hamiltonian, a task that can be hard to achieve near quantum
critical points. A powerful alternative is provided by the counterdiabatic 
modification of the Hamiltonian allowing for an arbitrarily quick implementation of
the adiabatic dynamics. Such a counterdiabatic driving protocol has been recently
proposed for the quantum Ising model [A. del Campo {\it et al.}, Phys. Rev. Lett. {\bf 109}, 115703 (2012)]. 
We derive an exact closed-form expression for all the coefficients  of the
counterdiabatic Ising Hamiltonian. 
We also discuss two approximations to the exact counterdiabatic Ising Hamiltonian quantifying 
their efficiency of the  dynamical preparation of the desired ground state. In
particular, these studies show how quantum criticality enhances finite-size effects 
in the counterdiabatic dynamics.  
\end{abstract}
%\pacs{05.30.Rt,05.50.+q}
\maketitle

\section{Introduction}
The ability to prepare a quantum system in its ground state is
crucial in adiabatic quantum computing \cite{FarhiSci2001},
quantum simulation of strongly correlated systems with trapped atoms and ions 
\cite{MaciekBook,Qsim}, studies of macroscopic superpositions of 
quantum phases in these systems
\cite{BDSciRep2012,MorigiPRA2011,MaschlerPRL2005}, etc. 

One way to experimentally  achieve this goal is to start from an
easy to prepare  ground state, and then  drive the system 
towards the desired ground state that is hard to prepare in a different way. 
This can be done through the change of the parameter of the Hamiltonian. 
If such driving is slow enough, the excitation of the system is typically exponentially small in
the driving rate \cite{Exp_adiabatic}. The adiabatic approximation tells us then
that the wave-function $|\psi(t)\ra$ during such evolution will be approximately given by 
\be
|\psi(t)\ra=\exp\B{-i\int_0^tdt'\varepsilon(t')-\int_{0}^{t}dt'\la GS(t')|\partial_{t'}GS(t')\ra}|GS(t)\ra,
\label{adiabatic}
\ee
where $|GS(t)\ra$ is the (instantaneous) ground state of the system's  Hamiltonian $\hat H_0[g(t)]$, 
$\hat H_0[g(t)]|GS(t)\ra=\varepsilon(t)|GS(t)\ra$, and $g(t)$ is the parameter driving  the
 evolution.

The main limitation of such a procedure comes from the requirement of the slow
change of the parameter of the Hamiltonian. The longer the evolution lasts,
the more affected by decoherence will be the evolved state.
This issue becomes especially troublesome in the quantum critical systems,
where the vanishing  energy gap near the critical point requires 
prohibitively slow time variations of the driving parameter
\cite{JacekAdv,PolkovnikovRMP2011}. 

One way to circumvent this problem is to temporarily modify the system's
Hamiltonian such that the adiabatic evolution (\ref{adiabatic}) is strictly 
imposed regardless of the rate of driving. To achieve that, one adds
a counterdiabatic term to the Hamiltonian \cite{DemirplakJPC2003,BerryJPA2009}
$$
\hat H_1=i g'(t) 
\sum_n\B{|\partial_g n(g)\ra\la n(g)|-\la n(g)|\partial_g n(g)\ra|n(g)\ra\la n(g)|}, \ \ g'(t)=\frac{dg}{dt},
$$
where $|n(g)\rangle$ are the  eigenstates of the system's Hamiltonian $\hat H_0[g]$. 
It is then easy to see that the wave-function (\ref{adiabatic}) satisfies
time-dependent Schr\"odinger equation
$$
i\frac{d}{dt}|\psi(t)\ra=\hat H[g(t)]|\psi(t)\ra,
$$
where the modified Hamiltonian of the system is 
$$
\hat H[g(t)]=\hat H_0[g(t)]+\hat H_1[g(t)].
$$
In fact, not only the ground state but also all the other eigenstates of $\hat H_0$ undergo
a perfect adiabatic evolution with the Hamiltonian $\hat H$ no matter how fast
the parameter $g(t)$ is changed. 
A comprehensive  discussion of different techniques allowing for a quick
preparation of the desired quantum state  is presented in a recent review \cite{AdolfoAdv2013};
see also Refs. \cite{AdolfoPRX2014,AdolfoRev}.
It is shown there that such techniques -- also known as shortcuts to
adiabaticity -- are applicable to both single-particle and many-body 
systems of  atoms, spins, etc.

\section{Model}
We consider the  Ising chain in the transverse field 
\be
\hat H_0 = -\sum_{i=1}^N \B{\sigma^x_i\sigma^x_{i+1} + g\sigma^z_i},
\label{H0}
\ee
where $\sigma_i^{x,y,z}$ denote the usual Pauli matrices acting on the $i$-th spin and $N$ is the 
number of spins in the chain.
We assume periodic boundary conditions and 
choose {\it even} $N\ge2$ for which the ground state lies in the positive parity
subspace \cite{BDfid4,ABQ2010}. We focus on magnetic fields $g\ge0$. When $g\in[0,1)$ the
system is in the ferromagnetic phase, while for $g\in(1,\infty]$ it is in the
paramagnetic phase. The critical point is at the magnetic field $g=1$. 
The quantum Ising model is exactly solvable. It is a prototypical model 
undergoing a quantum phase transition \cite{SachdevToday,Sachdev}. It serves as a test bed for
different theoretical concepts in many-body quantum physics.
Besides being of theoretical interest, such a model describes the magnetic
properties of  cobalt niobate CoNb$_2$O$_6$ \cite{ColdeaSci2010}. The number
of spins in such a system is thermodynamically large and avoids precise
experimental control.

The counterdiabatic term for the quantum Ising model 
has been found in Ref. \cite{AdolfoPRL2012}. It reads
\be
\hat H_1=-g'(t)\B{\sum_{m=1}^{N/2-1} h_m(g) H_1^{[m]} + \frac{1}{2}
h_{N/2}(g)H_1^{[N/2]}}, 
\label{H1}
\ee
where each term
$$ 
\hat H_1^{[m]}=\sum_{n=1}^N \left[\sigma_n^x\left(\prod_{j=n+1}^{n+m-1}\sigma_{j}^z\right)\sigma_{n+m}^y   
 + \sigma_n^y\left(\prod_{j=n+1}^{n+m-1}\sigma_{j}^z\right)\sigma_{n+m}^x \right],
$$
describes interactions of $m+1$ adjoint spins. 
Thus, to implement the counterdiabatic protocol one needs to be able to: (i) prepare 
an Ising chain with a predefined number of spins; (ii) 
implement the multi-spin interaction terms. The former should be 
available in the near future in cold ion \cite{MonroeAll,LanyonSci2011}, 
cold atom \cite{GreinerNat2011}, and NMR \cite{LiPRL2014} 
simulators of spin systems. 
The latter is even more demanding, but may  be possible with the techniques 
proposed in Ref. \cite{CasanovaPRL2012}.

The coefficients providing the strength of the $(m+1)$-body interaction terms are given 
by\footnote{
The $1/2$ prefactor in Eq. (\ref{hm_exact}) differs from the $1/4$ prefactor in 
the expression for $h_m(g)$ in Ref. \cite{AdolfoPRL2012} because we sum over positive 
momenta $k$ only, while the summation in Ref. \cite{AdolfoPRL2012} goes over positive and negative momenta,
i.e., $\pm k$ in our notation.} 
\be
h_m(g)=\frac{1}{2N}\sum_k \frac{\sin(k)\sin(mk)}{g^2-2g\cos(k)+1}, 
\label{hm_exact}
\ee
where both here and everywhere below 
$$
k=\frac{\pi}{N}, \frac{3\pi}{N},\cdots,\pi-\frac{\pi}{N}.
$$
It has been proposed in Ref. \cite{AdolfoPRL2012} that in the thermodynamic limit 
one can approximate the exact coefficients with
\be
\tilde h_m(g)=\frac{1}{8}\left\{
\begin{array}{ll}
g^{m-1} & {\rm for} \ 0\le g<1 \\
g^{-m-1} & {\rm for} \ g\ge1
\end{array}
\right..
\label{hm_approx}
\ee

This article is organized as follows. Section \ref{sec_counter}
is devoted to the derivation of the exact  closed-form expression for all 
the $h_m(g)$ coefficients and to the discussion of how they reflect the  symmetries 
of the quantum Ising model. 
Sections \ref{sec_tr} and \ref{sec_thermo} discuss two approximations to the exact counterdiabatic 
Hamiltonian. The former studies the dynamics induced by the interaction-truncated 
counterdiabatic Hamiltonian, while the latter focuses on the dynamics
resulting from the thermodynamic approximation (\ref{hm_approx}). 
In both sections, we quantify the efficiency of the dynamical  preparation of the  
desired  ground states. Our findings are briefly summarized in Sec. \ref{sec_discussion}.

\section{Exact counterdiabatic coefficients}
\label{sec_counter}

We will prove in this section the exact closed-form expression for  $h_m(g)$ coefficients,
the remarkably compact formula:
\be
h_m(g) = \frac{g^{2m}+g^N}{8g^{m+1}(1+g^N)} - \frac{\delta_{m0}}{8g},
\label{h_exact}
\ee
for $m=0,1,\cdots,N-1$, where the trivial $m=0$ case is added to simplify the
subsequent calculations. 

To prove it, we define one additional sum
\be
f_m(g) =\frac{1}{2N}\sum_k \frac{\cos(mk)}{\mn},
\label{fm_exact}
\ee
and conjecture  that 
\be
f_m(g)=\frac{1}{4g^m}\frac{g^N-g^{2m}}{(g^N+1)(g^2-1)}, 
\label{f_exact}
\ee
for $m=0,1,\cdots,N-1$. The reasoning leading to these two conjectures, 
Eqs. (\ref{h_exact}) and (\ref{f_exact}),  is explained in Appendix \ref{app_conj}.
We will  now simultaneously prove Eqs. (\ref{h_exact}) and (\ref{f_exact}) 
by  mathematical induction.

First, we stress that  Eqs. (\ref{h_exact}) and (\ref{f_exact}) 
are correct for even $N\ge2$ only. 
Moreover, it is worth to notice that all we need  for the counterdiabatic Hamiltonian 
is the expression for the $h_m(g)$ coefficients for $m\le N/2$, so we will prove  
a bit more  general result. Finally, we mention that although we assume $g>0$
in the proof, Eqs. (\ref{h_exact}) and (\ref{f_exact}) work for $g<0$ as
well. Indeed, from Eqs. (\ref{hm_exact}) and (\ref{fm_exact}) one gets that
$h_m(-g)=(-)^{m+1}h_m(g)$ and $f_m(-g)=(-)^mf_m(g)$. The same
$g\leftrightarrow-g$ mapping is
exhibited by Eqs. (\ref{h_exact}) and (\ref{f_exact}).

The induction starts from $m=0$, where we find that indeed Eqs.
(\ref{h_exact}) and (\ref{f_exact}) hold: The former holds trivially, 
while the latter  immediately follows  from the
following identity provided in Sec. 1.382 of Ref. \cite{Ryzhik}
\be
\sum_{k}\BB{\frac{\sin^2(k/2)}{\sinh(x)}+\frac{\tanh(x/2)}{2}}^{-1}=N\tanh\B{\frac{Nx}{2}},
\label{ryzhik}
\ee
after substituting $x=\ln(g)$.

Second, we use Eqs. (\ref{hm_exact}) and (\ref{fm_exact}) to show that 
\be
h_{m+1}(g) = \frac{g^2+1}{2g}h_m(g)-\BB{\frac{g^2-1}{2g}}^2f_m(g)+\frac{g^2+1}{16g^2}\delta_{m0}, \
\ f_{m+1}(g) = \frac{g^2+1}{2g}f_m(g) - h_m(g) - \frac{\delta_{m0}}{8g}. 
\label{recur}
\ee
From these relations it is clear why we have to consider the additional series,
$f_m(g)$, to inductively prove the conjectured expression for $h_m(g)$.
Eqs. (\ref{recur}) can be written in such a compact form with the help of the following
identities valid for integer $m\in[0,N-2]$
\begin{align}
\label{f1}
&\frac{1}{2N}\sum_k\frac{\cos(mk)\cos(k)}{\mn}=\frac{g^2+1}{2g}f_m(g)-\frac{\delta_{m0}}{8g}, \\
\label{f2}
&\frac{1}{2N}\sum_k\frac{\cos(mk)\sin^2(k)}{\mn}=-\BB{\frac{g^2-1}{2g}}^2f_m(g)+ 
\frac{g^2+1}{16g^2}\delta_{m0}+\frac{\delta_{m1}}{16g}, \\
\label{f3}
&\frac{1}{2N}\sum_k\frac{\sin(mk)\sin(k)\cos(k)}{\mn}=\frac{g^2+1}{2g}h_m(g)-\frac{\delta_{m1}}{16g}.
\end{align}
Eqs. (\ref{f1}) -- (\ref{f3}) are derived in Appendix \ref{app_sums}.

Third, substituting Eqs. (\ref{h_exact}) and (\ref{f_exact}) into Eqs. (\ref{recur}) one finds that
$$
h_{m+1}(g)= \frac{g^{2m+2}+g^N}{8g^{m+2}(1+g^N)}, \ \
f_{m+1}(g)=\frac{1}{4g^{m+1}}\frac{g^N-g^{2m+2}}{(g^N+1)(g^2-1)}, 
$$
which agrees with the conjectured expressions (\ref{h_exact}) and (\ref{f_exact}) taken at $m+1$. 
Such an inductive reasoning can be carried out for $m=0,1,\cdots,N-2$, for which
Eqs. (\ref{f1}) -- (\ref{f3}) hold. This concludes the inductive proof
that Eqs. (\ref{h_exact}) and (\ref{f_exact}) are valid for
$m=0,1,\cdots,N-1$. Note that the largest $m$ for which one can inductively 
prove Eq. (\ref{h_exact}) is 
indeed $N-1$, because  Eq. (\ref{h_exact}) gives clearly  wrong result for
$m=N$. We will now point out some general features of 
expression (\ref{h_exact}).

The $h_m(g)$ coefficients follow a simple  law 
under the Kramers-Wannier duality transformation 
(this symmetry of the quantum Ising model is discussed in Ref. \cite{Kramers})
$$
g \leftrightarrow \frac{1}{g}.
$$
Indeed, using Eq. (\ref{h_exact}) one easily finds that 
\be
g h_m(g) = \frac{1}{g} h_m\B{\frac{1}{g}}.
\label{duality}
\ee
This equation provides the ferromagnetic/paramagnetic duality of the $h_m(g)$ coefficients, i.e., 
a one-to-one mapping between their values  in the ferromagnetic and
paramagnetic phases.
Thus, Eq. (\ref{duality}) explicitly shows how the counterdiabatic
Hamiltonian reflects the symmetry of the system for which it is written down.
Moreover, we mention that duality symmetry (\ref{duality}) implies that the maximum of the $h_m(g)$ 
coefficients is not at the quantum critical point in any finite Ising chain (\ref{H0}).
This follows from the observation that one can rewrite Eq. (\ref{duality}) to
the form
\be
h_m(g)=\frac{\varphi_m(g)}{g}, \ \ \varphi_m(g)=\varphi_m\B{\frac{1}{g}}.
\label{phi}
\ee
Since $d\varphi_m(g)/dg=0$ at $g=1$, $dh_m(g)/dg$ cannot be equal to zero at the critical point. 
The $1/g$ factor in Eq. (\ref{phi}) shifts the maximum into the ferromagnetic phase, which can be 
directly verified from Eq. (\ref{h_exact}) as well.

A similar symmetry property have been found in the studies of the related
quantity, fidelity susceptibility providing  the overlap of two ``nearby'' ground states
\cite{BDfid3}. Just as $h_m(g)$, fidelity susceptibility is peaked near the critical
point \cite{GuReview,GuExp}. Besides the location of the critical point, it also encodes  the
universal critical exponent $\nu$ providing the scaling of the correlation length $\xi(g)$
with the distance from the critical point \cite{ABQ2010}. We can also see here the
influence  of quantum criticality  by noting that $h_m(g)$ depends on
$m$ and $N$ through the combinations
\be
g^m=\exp\B{\pm\frac{m}{\xi(g)}}, \ \ g^N=\exp\B{\pm\frac{N}{\xi(g)}},
\label{przez_xi}
\ee
where 
$$
\xi(g)=1/|\ln(g)|\approx |g-1|^{-1}
$$
is the correlation length of  the quantum Ising model \cite{BarouchPRA1971}; 
upper/lower sign corresponds to the paramagnetic/ferromagnetic phase. 
The first of these relations have been noticed  in Ref. \cite{AdolfoPRL2012}
as  it is properly captured by the thermodynamic approximation
(\ref{hm_approx}). Assuming that the divergence of the correlation length 
near the critical point $g_c$ is  given by the standard expression
\be
\xi(g)\sim |g-g_c|^{-\nu},
\label{xi}
\ee
one may speculate that the magnitude of
the $(m+1)$-body counterdiabatic interactions in other critical systems  will depend 
on the interaction range and the system size through the combinations $m|g-g_c|^\nu$ and 
$N|g-g_c|^\nu$. It would be interesting to verify this simple conjecture.

\section{Truncated counterdiabatic dynamics}
\label{sec_tr}
The main complications in the experimental implementation of the counterdiabatic
dynamics should come from the need to engineer the multi-spin
interactions. Suppose now that we can implement only the counterdiabatic terms 
coupling up to $M+1$ spins, i.e., we substitute 
\be
\bar h_m(g)=
\left\{
\begin{array}{ll}
\frac{g^{2m}+g^N}{8g^{m+1}(1+g^N)} & \ \ {\rm for} \  \ 1\le m \le M \\
0  & \ \ {\rm for} \ \ m > M 
\end{array}
\right.
\label{hm_tr}
\ee
for $h_m(g)$ in the Hamiltonian (\ref{H1}). If $M=N/2$ the full counterdiabatic protocol is
implemented and an exact preparation of the desired ground state takes place.
In the opposite limit, i.e. when $M=0$, the whole counterdiabatic Hamiltonian 
disappears and we are left with the system  susceptible to the 
non-adiabatic dynamics across the critical point \cite{DornerPRL2005,JacekPRL}. 
The purpose of this
section is to quantify the probability of finding the system in the
ground state by the end of time evolution, if the dynamics is driven by the 
truncated counterdiabatic Hamiltonian.
The study of the density of defects created due to the truncation of the 
 counterdiabatic Hamiltonian was performed in Ref.
\cite{AdolfoPRL2012}. Thus, our work provides a complementary insight into the 
influence of the truncation on the dynamics of the Ising chain. It is also
worth to mention that a different simplification of the counterdiabatic 
Ising Hamiltonian has been recently discussed in Ref. \cite{Saberi2014}.

The time-dependent modulation of the magnetic field is chosen such that 
at the beginning and the end of the time evolution -- $t=0$ and $t=T$,
respectively -- $\hat H=\hat H_0[g(t)]$. Moreover, we assume that the
magnetic field changes from $g_0$ to $g_f$. These requirements
impose the following boundary conditions
$$
g(0)=g_0,\ \ g'(0)=0,\ \ g(T)=g_f, \ \  g'(T)=0.
$$

They are satisfied, for example, by  a simple polynomial ansatz of the form
\be
g(t)=g_0+ 3\B{g_f-g_0}\B{\frac{t}{T}}^2 - 2\B{g_f-g_0}\B{\frac{t}{T}}^3,
\label{g_of_t}
\ee
which we will use below. Such an ansatz has been used in previous studies as
well \cite{Saberi2014}. 
We start the evolution from $g_0=5$,
i.e. far away from the critical point in the paramagnetic phase,
and evolve the system deeply into the ferromagnetic phase all the way to $g_f=0$.
This is the typical setup for the
studies of the dynamics of the quantum Ising chain (see e.g. Refs. \cite{JacekPRL,DornerPRL2005,AdolfoPRL2012}).

Although the experimental implementation of the counterdiabatic Hamiltonian
(\ref{H1})
is challenging due to the presence of the multi-spin interactions, the theoretical
description of the dynamics of the system driven by the non-local counterdiabatic
Hamiltonian is straightforward. This is seen by performing  the
Jordan-Wigner transformation: 
$$
\sigma^z_n=1-2\hat c_n^\dag\hat c_n, \ \ \sigma^x_n+i\sigma_n^y=2\hat c_n\prod_{l<n}\B{1-2\hat c_l^\dag\hat c_l},
$$
where $\hat c_n$ are fermionic operators, and then going to the  momentum space through the substitution 
$$
\hat c_n= \frac{\exp(-i\pi/4)}{\sqrt{N}}\sum_{K=\pm k} \hat c_K\exp(iKn).
$$
After these transformations one obtains
$$
\hat H = 2\sum_k \hat\psi_k^\dag 
\BB{\sin(k)\sigma^x - g'(t)q_k(g)\sigma^y-\B{\cos(k)-g}\sigma^z}
\hat\psi_k, \ \ 
\hat\psi_k = \B{
\begin{array}{c}
\hat c_k\\
\hat c_{-k}^\dag
\end{array}
},
$$
where 
\be
q_k(g)= 2\sum_{m=1}^{N/2-1}h_m(g)\sin(km)+h_{N/2}(g)\sin(kN/2).
\label{qk}
\ee
The evolution starts from a ground state of the Ising Hamiltonian, i.e., from a product state 
$$
\begin{aligned}
|GS(g)\rangle &= \prod_{k} 
 \B{\cos\B{\frac{\theta_k}{2}} -
\sin\B{\frac{\theta_k}{2}}\hat c_k^\dag\hat c_{-k}^\dag}|{\rm vac}\rangle, \\
\sin \theta_k &= \frac{\sin(k)}{\sqrt{g^2-2g\cos(k)+1}}, \ \ \cos\theta_k= \frac{g-\cos(k)}{\sqrt{g^2-2g\cos(k)+1}}
\end{aligned}
$$
taken at $g=g_0$. The $|{\rm vac}\rangle$ state is annihilated by all $\hat c_k$ operators.

\begin{figure}[t]
\includegraphics[width=9.369cm]{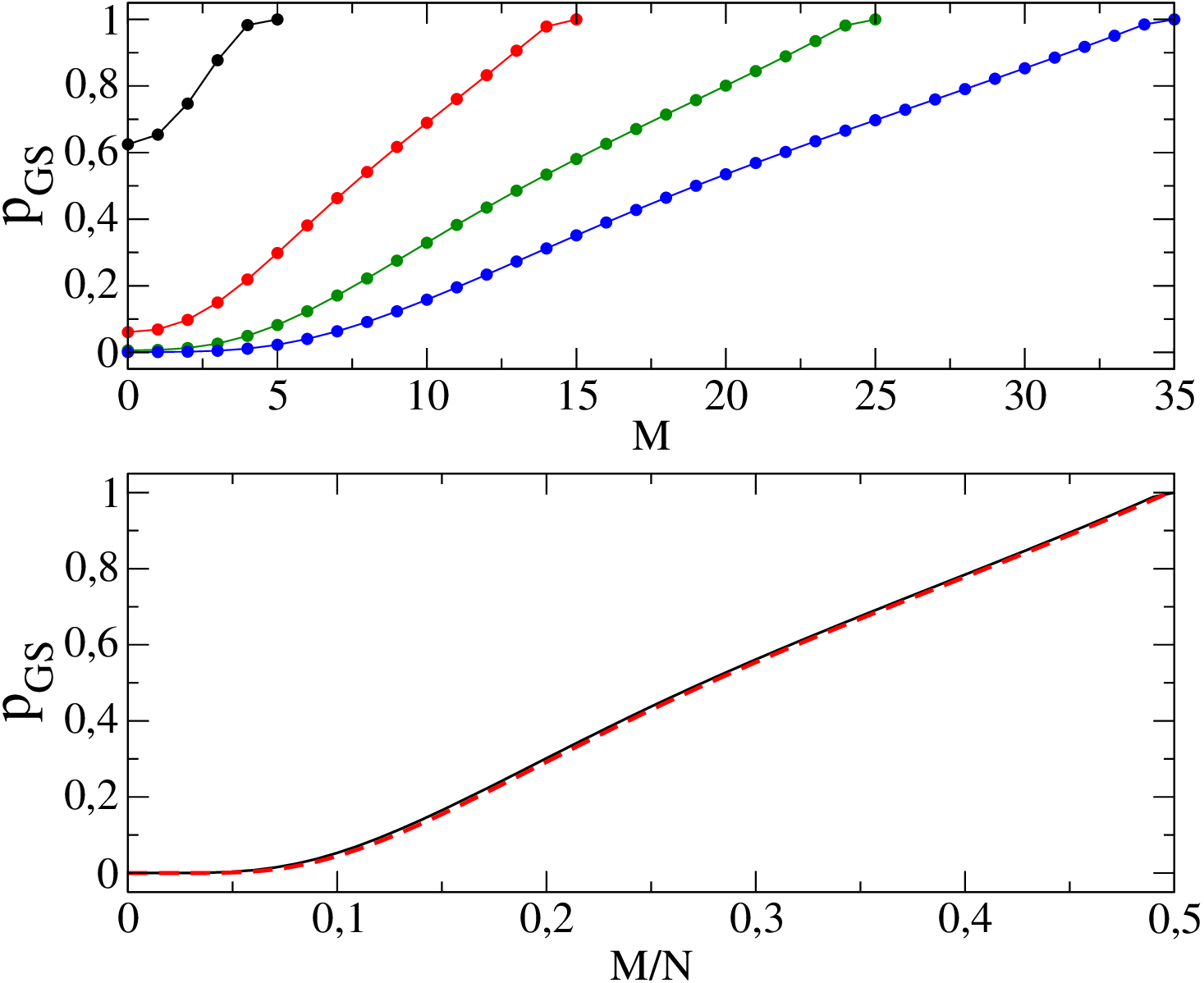}
\caption{
Probability of finding the system in the ground state by the end of
time evolution as a function of the range of counterdiabatic interactions
(\ref{hm_tr}).
Upper panel: lines from left to right correspond to 
 $N$ equal to $10$ (black), $30$ (red), $50$ (green), $70$
(blue). Lower panel: solid black line corresponds to $N=100$ while the red
dashed line to $N=200$. The evolutions are driven by the time-dependent 
magnetic field (\ref{g_of_t}), where $g_0=5$, $g_f=0$, and $T=10$.
}
\label{fig_range}
\end{figure}

Solving the Schr\"odinger equation $i\frac{d}{dt}|\psi(t)\rangle=\hat H[g(t)]|\psi(t)\rangle$
with the initial wave-function $|\psi(t=0)\ra=|GS(g_0)\rangle$,  one finds
that 
$$
\begin{aligned}
&|\psi(t)\rangle =\prod_{k} \B{u_k(t) - v_k(t)\hat c_k^\dag\hat c_{-k}^\dag}|{\rm vac}\rangle,\\
&i\frac{d}{dt}
\left(
\begin{array}{c}
v_k \\
u_k
\end{array}
\right)
=2\left(
\begin{array}{cc}
g-\cos(k) & -\sin(k) -ig'(t)q_k\\
-\sin(k)+ig'(t)q_k  & \cos(k)-g
\end{array}
\right)
\left(
\begin{array}{c}
v_k \\
u_k
\end{array}
\right).
\end{aligned}
$$

If we would substitute the exact $h_m(g)$ into Eq. (\ref{qk}), we would get 
\be
q_k(g) = \frac{1}{4}\frac{\sin(k)}{g^2-2g\cos(k)+1},
\label{qk_ex}
\ee
which guarantees the perfect preparation of the ground state by the end of time
evolution \cite{AdolfoPRL2012}. Namely, 
$$
p_{GS}= |\la\psi(T)|GS(g_f)\ra|^2
$$
equals unity in such a case.  What happens when we use coefficients (\ref{hm_tr}) instead of (\ref{h_exact}) 
is plotted in Fig.
\ref{fig_range}. 
Several remarks are in order now.

First, when $M=N/2$, no truncation is involved, the perfect preparation of the ground
state happens. Therefore,  $p_{GS}$ should be equal to one. Our numerical simulations 
provide $|p_{GS}-1| = {\rm O}\B{10^{-12}}$ 
for all the $M=N/2$ cases illustrated  in Fig. \ref{fig_range}. This 
shows that the  numerical errors are negligible in our computations.

Second, we consider the $M=0$ case, when  counterdiabatic dynamics is
absent. We see that the probability of finding the system in the ground
state  quickly goes to zero as the system size increases. The need to 
counteract this tendency motivates the introduction of the counterdiabatic
driving protocol that we discuss. It should be also said that this
disappearance of $p_{GS}$ comes from the fact that the gap in the excitation spectrum 
scales as $1/N$ near the critical point. As a result, bigger systems get more excited due to the
dynamical crossing of the critical point. The excitations that
are created in the course of such dynamics are known as kinks  (see e.g. Ref. \cite{DornerPRL2005,JacekPRL}).

Third, for large enough systems --  $N$ larger than a few tens for the
evolutions from Fig. \ref{fig_range} --
we find that $p_{GS}$ depends mostly on the  $m/N$  ratio (see the lower panel).
Using Fig. \ref{fig_range}, we can correlate the range of the counterdiabatic
interactions with the probability of ground state preparation. 
On the one hand, we notice that even substantially  truncated counterdiabatic  Hamiltonian
greatly increases $p_{GS}$ with respect to what one ends up with when no
counterdiabatic driving is present. On the other hand, to
prepare the desired ground state with probability larger than $95\%$, 
virtually all the terms in the counterdiabatic 
Hamiltonian have to be present. This observation quantifies how demanding is 
the  counterdiabatic dynamics if high-probability of ground state preparation  is sought for.

\section{Thermodynamic approximation}
\label{sec_thermo}
We will discuss here   the thermodynamic approximation (\ref{hm_approx})
to the counterdiabatic Hamiltonian to illustrate  the interplay between 
finite-size effects and quantum criticality in the counterdiabatic dynamics. 

\begin{figure}[t]
\includegraphics[width=9.369cm,clip=true]{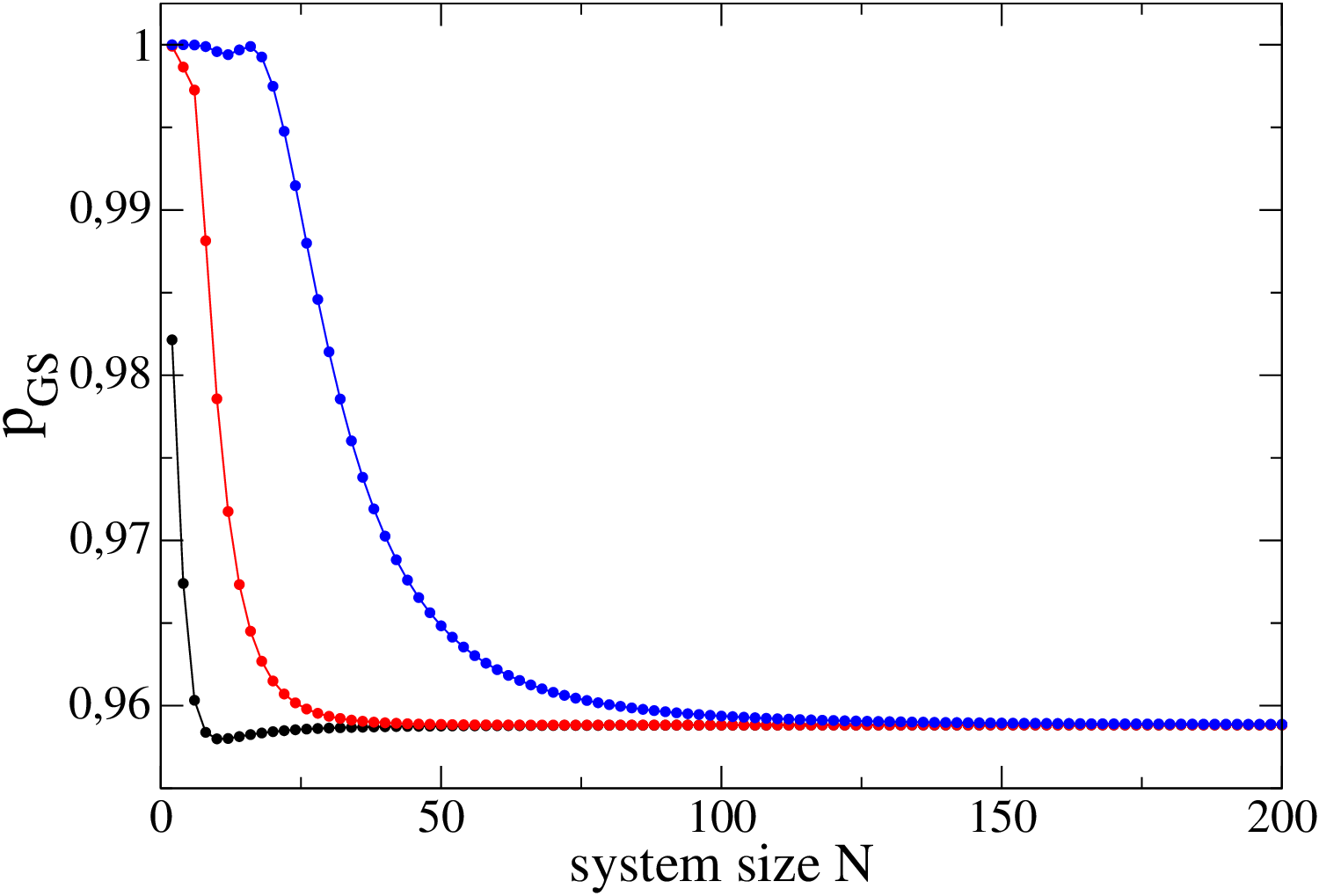}
\caption{
Probability of finding the system in the ground state by the end of
time evolution as a function of the system size.
Thermodynamically-approximated coefficients (\ref{hm_approx}) are used in the
counterdiabatic Hamiltonian (\ref{H1}).
The evolutions are driven by the time-dependent magnetic field (\ref{g_of_t}),
where  $g_0=5$ and $g_f=0$. The  black, red, and blue  lines 
-- bottom to top -- correspond to 
the evolution times $T$ equal to $1$, $10$, and $100$, respectively. 
}
\label{termo_prob}
\end{figure}

The numerics is performed in the same way as in the previous section, except 
we use the coefficients $\tilde h_m(g)$ instead of $\bar h_m(g)$.
The system-size dependence of the probability of ground state preparation  $p_{GS}$ is
illustrated in Fig. \ref{termo_prob}.
This figure shows that the lack of finite-size effects in the 
counterdiabatic coefficients limits the maximum efficiency of the preparation of 
the ground state.

On the other hand, the efficiency of the driving with the
thermodynamic Hamiltonian is quite substantial and equals about $96\%$
in the large system  limit (Fig. \ref{termo_prob}). This can  be
compared to the truncation approximation studied in Sec. \ref{sec_tr}.
To achieve the same probability of the ground state preparation with the latter 
approach, one needs to perfectly emulate all but one or two longest-range 
interaction terms (Fig.
\ref{fig_range}).

The fact that $p_{GS}$ obtained through the thermodynamic approximation 
does not approach unity as the system size increases is a bit surprising.
In fact, it has an interesting explanation that we will provide below.

\begin{figure}[t]
\includegraphics[width=9.369cm]{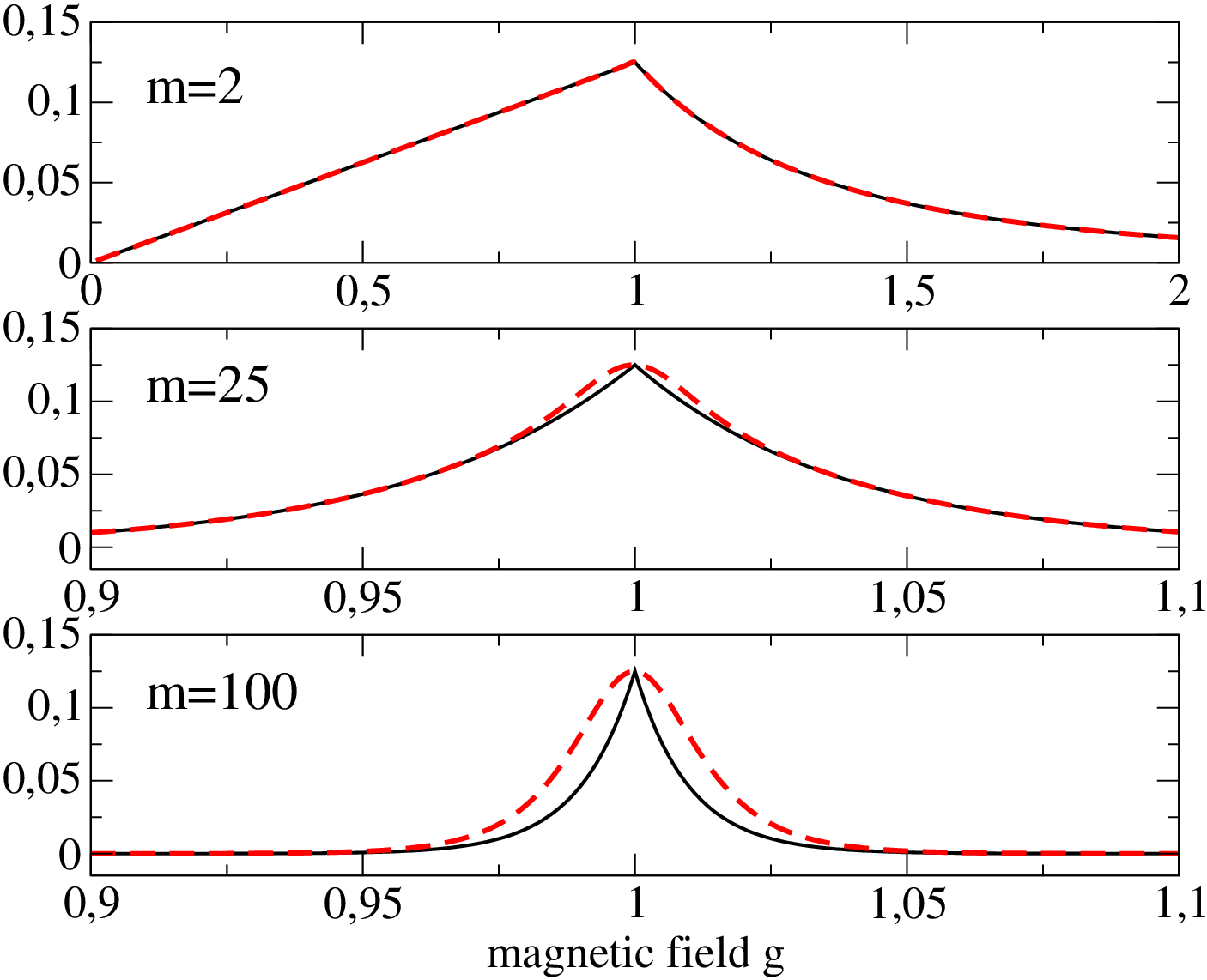}
\caption{Exact $h_m(g)$ vs. approximate $\tilde h_m(g)$ coefficients. 
Exact results (red dashed line)  come from Eq. (\ref{h_exact}), 
while the thermodynamic approximation (black solid line) is provided by Eq. (\ref{hm_approx}). 
The number of spins $N=200$. 
}
\label{fig_hm_large}
\end{figure}

The thermodynamic approximation (\ref{hm_approx}) follows from the replacement of
the sum by the integral in Eq. (\ref{hm_exact})
$$
\sum_k \to \frac{N}{2\pi}\int_0^\pi dk.
$$
We expect that such an approximation will work well when $m/N\ll1$ and that 
it will fail  for $m$ of 
the order of $N/2$. This expectation is based on the observation that 
when $k\to k+2\pi/N$, the summand in Eq. (\ref{hm_exact}) changes slowly
in the former case and quickly in the latter one.
Since there are  simple  analytical expressions for the 
thermodynamic (\ref{hm_approx}) and  exact (\ref{h_exact}) 
$h_m(g)$ coefficients, the reader  can easily  verify these expectations. We 
illustrate these remarks  in Fig. \ref{fig_hm_large}.

Next, we substitute the $\tilde h_m(g)$ coefficients into Eq. (\ref{qk})
to see how the  inaccuracies of the thermodynamic approximation affect time
evolution. We get 
\be
q_k(g)=\frac{1}{4}\frac{\sin(k)}{g^2-2g\cos(k)+1}\pm
\frac{g^{\pm N/2}}{8g} \frac{g^2-1}{g^2-2g\cos(k)+1}\sin(kN/2),
\label{qk_th}
\ee
where the upper/lower sign should be used in the ferromagnetic/paramagnetic
phase. The first term in this expression reproduces
the exact form of the $q_k(g)$ factor (\ref{qk_ex}) needed for a perfect 
adiabatic preparation of any eigenstate of the Ising Hamiltonian (\ref{H0}).
The second one is the
correction resulting from the above-described deficiencies of the thermodynamic approximation. 
Rewriting $g^{\pm N/2}$ as $\exp\B{-N/2\xi(g)}$, we see that this
correction plays a role only in the quantum critical regime, 
where the correlation length $\xi(g)$ is comparable or greater than the system
size $N$ [note that this is possible as $\xi(g)$ stands for the 
correlation length of the infinite system (\ref{xi})].  

\begin{figure}[t]
\includegraphics[width=9.369cm]{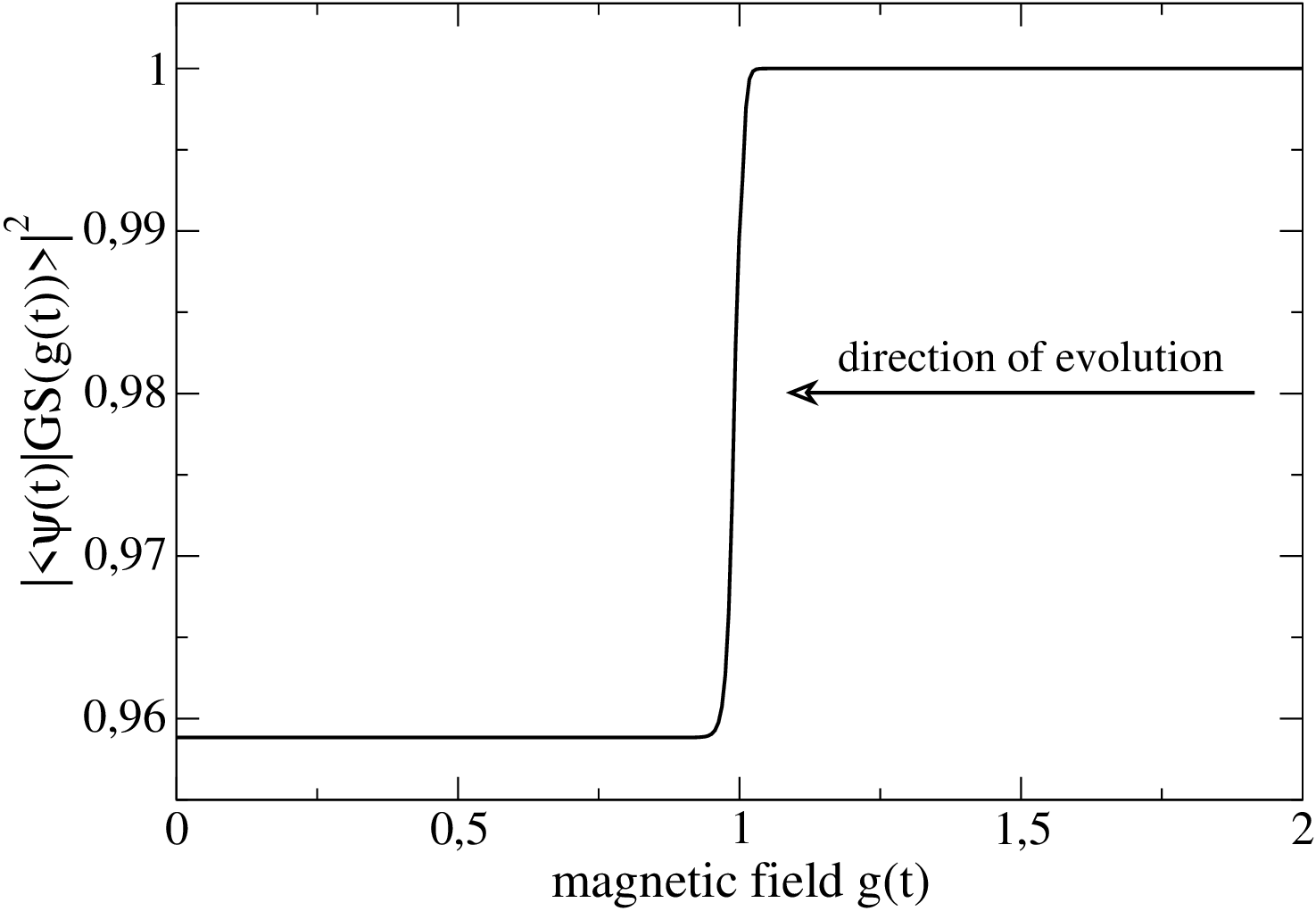}
\caption{Probability of finding the system in the instanteneous ground state 
during time evolution from the paramagnetic to the ferromagnetic phase.
Thermodynamically-approximated coefficients (\ref{hm_approx}) are used in the
counterdiabatic Hamiltonian (\ref{H1}).
The evolution is  driven by the time-dependent 
magnetic field (\ref{g_of_t}), where $g_0=5$, $g_f=0$, and $T=10$.
The system size  $N=200$. 
Compare this figure to Figs. \ref{termo_prob} and \ref{fig_hm_large}.
}
\label{fig4}
\end{figure}

These remarks give us the following picture of the dynamics driven by the
thermodynamically-approximated counterdiabatic Hamiltonian. Suppose the 
evolution starts far away from the critical point 
 in the paramagnetic phase. The population of the instanteneous eigenstates
 of the time-dependent Ising
 Hamiltonian (\ref{H0}) does not change until we reach the neighborhood of the critical 
 point, where $|g-1|=O(1/N)$. For such magnetic fields the second term in Eq.
 (\ref{qk_th}) excites the system. 
 Soon after leaving the neighborhood of the critical point, the adiabatic dynamics
 resumes and the excited system is  driven without 
the  change of its  population in the  instanteneous eigenbasis.
This is illustrated in Fig. \ref{fig4}.

\section{Discussion}
\label{sec_discussion}

We have found the exact-closed form expression 
for all the coefficients of the counterdiabatic Ising  Hamiltonian (\ref{h_exact}).
Our results are very compact and provide the ultimate simplification of this Hamiltonian. 
We have  also discussed  how  the Kramers-Wannier duality  and quantum
critical properties  of the Ising chain are imprinted onto the 
counterdiabatic Hamiltonian (\ref{duality},\ref{przez_xi}). 
It should be also stressed that our Eqs. 
(\ref{h_exact}) and (\ref{f_exact}) provide neat summation formulas 
that can be used in other contexts as well.

The main complications in the experimental implementation of the counterdiabatic Hamiltonian
come from its reliance on the  many-body interactions that are hard-to-induce
in  real physical systems. We have  truncated the exact counterdiabatic Hamiltonian removing 
its  longest-range interaction terms. Using such a Hamiltonian, we have
quantified the efficiency of the preparation of the desired ground state. We
have shown that the truncated Hamiltonian can significantly improve the
probability  of
finding the system in the ground state by the end of time evolution.
However, if the goal is to prepare the ground state with a substantial 
probability, say greater than $95\%$, then this approximation becomes 
inefficient as virtually all the 
multi-spin interaction terms have to be kept  to
achieve such a goal. 

We have also considered a simple thermodynamic approximation to the counterdiabatic 
Hamiltonian coefficients and explained its limited efficiency in the dynamical 
preparation of the desired ground state.
It should be noted that there is no  incentive to actually use that approximation 
given the compactness of our exact result (\ref{h_exact}). We think, however, that 
the study of this approximation provides some insights into the role of the 
finite-size effects in the counterdiabatic dynamics. It shows that the 
thermodynamic approximation leads to the undesired excitation of the system 
in the quantum critical regime, where finite-size effects cannot be overlooked 
if one aims at high-probability preparation of the target ground state.

\section{Acknowledgments}
This work has been supported by the Polish National
Science Centre (NCN) grant DEC-2013/09/B/ST3/00239.
I would like to thank Adolfo del Campo for several useful suggestions and
stimulating discussions.

\appendix
\section{Conjecture}
\label{app_conj}

We describe below how one can conjecture fairly complicated Eqs. (\ref{h_exact}) and
(\ref{f_exact}). 
The main difficulty here is that Eqs. (\ref{hm_exact}) and (\ref{fm_exact})  depend on both 
the system size $N$ and the range of counterdiabatic  interactions  $m$. 
In the following, we  factor out the $N$-dependence
exactly, leaving the $m$-dependence for  an easy educated guess.
We focus on showing the key steps in this calculation leaving the details to the reader.

First, we notice that the denominator in Eq. (\ref{hm_exact}) can be rewritten
as 
\be
g^2-2g\cos(k)+1 = 4g\BB{\sin^2(k/2)+\sinh^2(x/2)},
\label{xparam}
\ee
where $x=\ln(g)$. The calculations can be  more efficiently
performed in the $x$-parameterization.

Second, we derive the recurrence relation for 
$$
w_{n}=\sum_k\frac{\sin^{n}(k/2)}{\sin^2(k/2)+\sinh^2(x/2)},
$$
which reads 
\be
w_{2n+2}=\frac{N}{2^{2n+1}}\binom{2n}{n}-w_{2n}\sinh^2(x/2), \ \ n=0,1,\cdots,N-1.
\label{rec}
\ee
To solve this relation, we find from identity (\ref{ryzhik}) that 
\be
w_0(x) = N\frac{\tanh(Nx/2)}{\sinh(x)}.
\label{w0}
\ee
Note that this equation is valid for even $N\ge2$, which we assume throughout
this article.
Combining Eqs. (\ref{rec}) and (\ref{w0}) we get 
\be
\begin{aligned}
\sum_k \frac{\sin^{2n}(k/2)}{\sin^2(k/2)+\sinh^2(x/2)}=
N(-)^n\sinh^{2n}(x/2)\frac{\tanh(Nx/2)}{\sinh(x)}+
N\sum_{s=0}^{n-1}
\binom{2s}{s}\frac{(-)^{n-s-1}}{2^{2s+1}}\sinh^{2n-2s-2}(x/2),
\end{aligned}
\label{sum_m}
\ee
which  is valid for $n=0,1,\cdots,N$ 
(the sum on the right-hand side yields zero for $n=0$).

Third, we expand numerator in Eq. (\ref{hm_exact}). We obtain 
\be
\sin(k)\sin(mk) =\sum_{s=0}^{m} {\cal A}^m_s \sin^{2s+2}(k/2), \ \
{\cal A}^m_s=(-)^s2^{2s+1}\frac{(m+s-1)!}{(m-s)!(2s+1)!}\B{2m^2+s}
\label{Asin}
\ee
for $m>0$ (${\cal A}^0_0=0$).
Taking $x=\ln(g)$ and combining Eqs. (\ref{sum_m}) and (\ref{Asin}),  we
find that 
\be
\begin{aligned}
h_m(g) = \frac{1}{8g}\sum_{j=0}^m(-)^j{\cal A}^m_j
\BB{\frac{(g-1)^2}{4g}}^j
\B{ \frac{g^N+g}{(g+1)(g^N+1)} +
\sum_{s=1}^j
\binom{2s}{s}\frac{(-)^s}{2^{2s+1}}\BB{\frac{4g}{(g-1)^2}}^s}
\end{aligned}
\label{ytrewq}
\ee
for $m=1,2,\cdots,N-1$.
This equation is fairly complicated and we do not know how to simplify it. It
has, however, the $N$-dependence in a closed-form, which greatly
facilitates the following considerations. 
Substituting $m=1,2,\cdots$ into Eq. (\ref{ytrewq}) one easily guesses that the general
expression for $h_m(g)$ is provided by Eq. (\ref{h_exact}). 
It should be stressed that it would be very difficult to conjecture Eq.
(\ref{h_exact})  right from Eq. (\ref{hm_exact}).

Fourth, we expand the numerator  in Eq. (\ref{fm_exact}) getting
\be
\cos(mk) =\sum_{s=0}^{m} {\cal B}^m_s \sin^{2s}(k/2), \ \
{\cal B}^m_s=\frac{(-)^s2^{2s}m(m+s-1)!}{(m-s)!(2s)!}
\label{cos_mk}
\ee
for $m>0$ (${\cal B}^0_0=1$). Taking again $x=\ln(g)$ and this time combining Eqs. (\ref{sum_m}) and (\ref{cos_mk}), we 
get 
$$
\begin{aligned}
f_m(g) = \frac{1}{8g}\sum_{j=0}^m(-)^j{\cal B}^m_j
\BB{\frac{(g-1)^2}{4g}}^j
\B{\frac{2g}{g^2-1} \frac{g^N-1}{g^N+1} -
\sum_{s=0}^{j-1}
\binom{2s}{s}\frac{(-)^s}{2^{2s+1}}\BB{\frac{4g}{(g-1)^2}}^{s+1}
   }
\end{aligned}
$$
for $m=1,2,\cdots,N-1$. As before, substituting $m=1,2,\cdots$ 
we quickly find that Eq. (\ref{f_exact}) is a good candidate for an exact
expression for series (\ref{fm_exact}).

\section{Auxiliary sums}
\label{app_sums}
{\it Derivation of Eq. (\ref{f1}).} We employ  the $x$-parameterization from
Eq. (\ref{xparam}).  
After elementary manipulations we find that 
$$
\frac{d}{dz}\BB{\sinh^2(z/2)\sum_k \frac{\cos(mk)}{\sin^2(k/2)+\sinh^2(z/2)}}=-\frac{1}{2}\frac{d}{dz}
\sum_k \frac{\cos(mk)\BB{1-\cos(k)}}{\sin^2(k/2)+\sinh^2(z/2)}.
$$
Integrating it over $z$ from $0$ to $x$ we get
$$
\sum_k\frac{\cos(mk)\cos(k)}{\sin^2(k/2)+\sinh^2(x/2)}=
\cosh(x)\sum_k \frac{\cos(mk)}{\sin^2(k/2)+\sinh^2(x/2)} -N\Delta_m, \ \
\Delta_m=
\left\{\begin{array}{ll}
0 & {\rm for} \ \ \frac{m}{N} \notin \mathbb{Z}\\
(-1)^{m/N} & {\rm for} \ \ \frac{m}{N} \in \mathbb{Z}
\end{array}
\right., 
$$
where $N\Delta_m$ comes from the evaluation of $2\sum_k\cos(mk)$ and
$\mathbb{Z}=0,\pm1,\pm2,\cdots$.
Substituting $x=\ln(g)$ and restricting the range of $m$ to $[0,N-2]$ we obtain Eq. (\ref{f1}).

{\it Derivation of Eq. (\ref{f2}).} Just as above, one  finds that 
$$
\frac{d}{dz}\BB{\sinh^2(z/2)\sum_k \frac{\cos(mk)\cos(k)}{\sin^2(k/2)+\sinh^2(z/2)}}=-\frac{1}{2}\frac{d}{dz}
\sum_k \frac{\cos(mk)\cos(k)\BB{1-\cos(k)}}{\sin^2(k/2)+\sinh^2(z/2)}.
$$
Repeating the same steps as above, we get
$$
\sum_k\frac{\cos(mk)\sin^2(k)}{\sin^2(k/2)+\sinh^2(x/2)}=
-\sinh^2(x)\sum_k \frac{\cos(mk)}{\sin^2(k/2)+\sinh^2(x/2)} +
N\cosh(x)\Delta_m + \frac{N}{2}\B{\Delta_{m-1}+\Delta_{m+1}},
$$
which is equivalent to Eq. (\ref{f2}) for $x=\ln(g)$ and integer $m\in[0,N-2]$.

{\it Derivation of Eq. (\ref{f3}).} This time we start from  
$$
\frac{d}{dz}\BB{\sinh^2(z/2)\sum_k \frac{\sin(k)\sin(mk)}{\sin^2(k/2)+\sinh^2(z/2)}}=-\frac{1}{2}\frac{d}{dz}
\sum_k \frac{\sin(k)\sin(mk)\BB{1-\cos(k)}}{\sin^2(k/2)+\sinh^2(z/2)}
$$
and end up with
$$
\sum_k\frac{\sin(k)\sin(mk)\cos(k)}{\sin^2(k/2)+\sinh^2(x/2)}=
\cosh(x)\sum_k \frac{\sin(k)\sin(mk)}{\sin^2(k/2)+\sinh^2(x/2)} -\frac{N}{2}\B{\Delta_{m-1}-\Delta_{m+1}},
$$
which can be transformed into Eq. (\ref{f3}) by  taking $x=\ln(g)$ and integer $m\in[0,N-2]$.

%\bibliography{reference} 

\end{document}